# A novel 3D display based on micro-volumetric scanning and real time reconstruction of holograms principle

Guangjun Wang[a,b]

[a]BOWEI INTEGRATED CIRCUITS CO.,LTD., The 13th Research Institute of China Electronics Technology Group Corporation, Shijiazhuang, Hebei 050200, P.R. China
[b] Key Laboratory of Physics and Technology for Advanced Batteries, Ministry of Education, College of Physics, State Key Laboratory of Superhard Materials,Jilin University, Changchun, 130012, PR China

**Abstract:**

Nowadays, the 3D display technology has attracted great academic and industrial attention due to its rapid development and applications in providing more realistic, natural, and extra depth images far superior than the traditional 2D display. But traditional 3D techniques suffer from many drawbacks and hard to meet the requirement of commercialization. Here, we designed a novel 3D display system which combined the advantage of volumetric 3D display and hologram technology. The novel display contains a micro-volumetric scanning system (MVS) and a real time reconstruction hologram system (RTRH). The micro-volumetric scanning system has similar structure of a micro-projector but it doesn't have the screen and its focal plane is scanning/moving back- and -forth constantly. Thus it can project a series of 2D images at different depth of field, which can form a 3D image in true 3D space (but the 3D image is a divergent one and can't be seen). The RTRH is an improvement of traditional hologram system. The most important feature of the RTRH is that the object (in traditional hologram system) is replaced by a virtual object-the projected 3D image (by the MVS). And the interference pattern of virtual object light and reference light is displayed on a rear projection screen. Thus, instead of illuminating the interference pattern recorded on a film (or generated on a LCD), it directly shows the lighted interference pattern (of object light and reference light) on a transparent projection screen. Then the hologram of the virtual object is reconstructed timely.

## 1. INTRODUCTION

Nowadays, the 3D display technology has attracted great academic and industrial attention due to its rapid development and applications in providing more realistic, natural, and extra depth images far superior than the traditional 2D display[1]. To achieve natural 3D visual perception, a considerable amount of efforts have been dedicated, and some practical designs have been proposed. But these traditional 3D techniques suffer from many drawbacks and hard to meet the requirement of commercialization. Among them the 3D display technology based on two eyes parallax is the most mature one[2]. But this kind of 3D displays easily cause visual fatigue because of the accommodation–vergence conflict [3, 4] besides some other disadvantages, i.e., the fixed range of the viewing distance, view position, and the fixed interpupillary distance value (usually 65mm). Also, the parallax barrier technology seriously reduces the brightness of images, so it is replaced by lenticular sheet technology in most products. In the lenticular based autostereoscopic display, the role of the lenticular lens is to magnify and transfer the information of specific pixels to a designated position[5]. However, the depth-reversal is considered as an inherent disadvantage of the lenticular based or the parallax barrier based autostereoscopic displays. Integral imaging is a kind of true 3D technology based on the principle of reversibility of light[6, 7]. But due to the poor resolution and complex structure, even it has been proposed by Lippmann for around a century, it has not got wide acceptance. Volumetric 3D display is another kind of true 3D displays, which has also some inherent limitations[8]. The mechanical rotating components limits the spatial of image's and reduces the resolution at the center of the display. Holography clearly distinguishes itself from all the above-mentioned techniques. Indeed, it has the ability to reconstruct the wavefront of the light scattered by an object and thus reproduce the sensation of a viewer has standing in front of a real object[9]. However, the lack of sufficient computational power to produce realistic computer-generated holograms and the absence of large-area and dynamically updatable holographic recording media have prevented realization of the concept. Meanwhile, it's difficult for computer-generated holograms method to display color images. In traditional hologram technology the recorded interferogram is used to set the "boundary condition" to get a static 3D image[9]. In order to display a dynamic 3D image a dynamic "boundary condition" is needed. In a typical computer-generated holograms system, spatial light modulator (SLM) is usually used to generate the dynamic patterns [10]. But this kind of hologram display can not display large and clear 3D image due to the limitations of SLM[4]. A delicate design, proposed by Takaki, overcomes this defect to some extent[11]. In order to get a large computer generated hologram, a horizontal scanner is introduced to scan the elementary hologram generated by the anamorphic imaging system. However, the default shortcoming of computer-generated holograms is that it is difficult to provide a SLM with the pixel counts high enough. A further complexity is to provide the data in realtime.

It is difficult to construct a three-dimensional image in free space allowing view from all directions or positions. In fact, there is no need to achieve a full view 3D image display. Audiences usually sit or stand at one side of a 3D display, so they don't care about other sides of 3D images that can't be seen at their position. From a mathematical point-of-view, the construction of 3D image in free space is a definite boundary solution problem of passive space. If the boundary condition (wavefront) is reconstructed the light field in the passive space will also be reconstructed. Based on these considerations, a new approach to display hologram was proposed by R. Häusslerur[12]. Instead of reconstructing the image that can be seen from a large viewing region, the primary goal is to reconstruct the wavefront that can only be seen at a viewing window (VW). This holographic display omits unnecessary wavefront information and significantly reduces the requirements on the resolution of the spatial light modulator and the computation effort compared to conventional holographic displays. But it still can't meet the requirement of commercialization. Here, we designed a novel 3D display system which combined the advantage of volumetric 3D display and hologram technology. The novel display contains a MVS and a RTRH. The micro-volumetric scanning system has similar structure of micro-projector but it doesn't have the screen and its focal plane is scanning/moving back- and -forth constantly. Thus it can project a series of 2D images at different depth of field, which can form a 3D image in true 3D space (but the 3D image is a divergent one and can't be seen). The RTRH is an improvement of traditional hologram system. The most important feature of the RTRH is that it use the projected 3D image (by the MVS) as a virtual object and the interference pattern of virtual object light and reference light is displayed on a transparent projection screen. Thus, instead of illuminating the interference pattern recorded on a film (or generated on a LCD), it directly shows the lighted interference pattern (of object light and reference light) on a transparent projection screen. Then the hologram of the virtual object is reconstructed timely.

## 2. PRINCIPLE AND DESIGN CONSIDERATION

Since the volumetric 3D display can display volumetric 3D images in true 3D space, we decide to use this method to realize the purpose of displaying 3D image. But traditional volumetric 3D displays usually need large scanning parts or volumetric media to display "voxel". These features make it difficult to achieve a practical 3D display. Thus, we designed a novel MVS. Rather than using a large movement part, we introduce a micro-scanning part and an optical magnifies part. Fig. 1. depicts the schematic of MVS. The configuration of MVS looks like a projector, which contains a display chip and a projector lens. What different is that the display chip (or lens) is scanning back and forth to generate a micro 3D image (inside the scanning space). After been project out (through the lens), the micro 3D image is been magnified (the viewing space). But the projected 3D image can't be seen, because it is a divergent one. With the help of RTRH, the divergent 3D image can be converting to a convergent one, and then it can be seen.

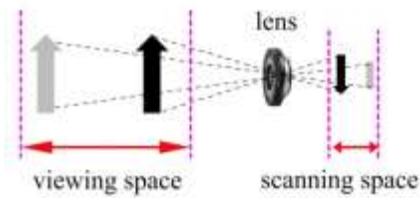

Fig. 1. Schematic diagram of MVS system

The working principle of RTRH is shown in Fig. 2 and Fig. 3. The MCV can project a virtual 3D object (image) in free space. Then the light from the projected virtual 3D object interference with the reference light and form an interferogram, which is the hologram of the virtual 3D object, on the screen.

The light emitted from the virtual 3D object can be mathematically expressed as:

$$O(x,y) = A_o(x,y)e^{i\varphi_o(x,y)} \qquad (1)$$

And for the reference light:

$$R(x,y) = A_r(x,y)e^{i\varphi_r(x,y)} \qquad (2)$$

Then the amplitude and intensity of interferogram on the screen satisfies:

$$A(x,y) = O(x,y) + R(x,y) \qquad (3)$$

$$I(x,y) = A(x,y) \cdot A^*(x,y) = O^2 + R^2 + 2A_rA_o\cos(\varphi_r - \varphi_o) \tag{4}$$

Where * denotes complex conjugate. For simplicity, the $\varphi_r(x,y)$ can be chosen as 0 and this is easy to implement under actual conditions.

It should be note that only the bright fringe allows light to pass through the screen. So it seems that the interferogram on the screen can modulate the light. Then at the modulating function of the interferogram the transmitted part of reference light can be written as:

$$R_t = (T + \beta \cdot I(x,y)) \cdot R(x,y)$$
$$= (T + \beta \cdot (O^2 + R^2))R(x,y) + \beta \cdot A_r^2 \cdot O^*(x,y) + \beta \cdot A_r^2 \cdot O(x,y) \tag{5}$$

Where both T and $\beta$ are constants. At proper condition, T becomes negligibly small. In the equation 5, the first term represents the attenuated reference light while the second term is the complex conjugate of $O(x,y)$, which means that it will generate a real image (of the virtual 3D object). Take the conjugate relationship between the virtual 3D object and the real image into consideration, this real image is a convergent one and can be seen by eyes at the conjugate position of the lens(of projector), as shown in Fig. 3. The third term has the same form of $O(x,y)$, which will produce a virtual image. The difference between the real image and virtual image is that the former is a converged image while the latter is an emanative one. Thus the light from the real image converges to a certain position (the conjugate position of the lens) and can be seen completely at this position. But the light from the virtual one diffuses to different directions so it can't be seen completely at any position.

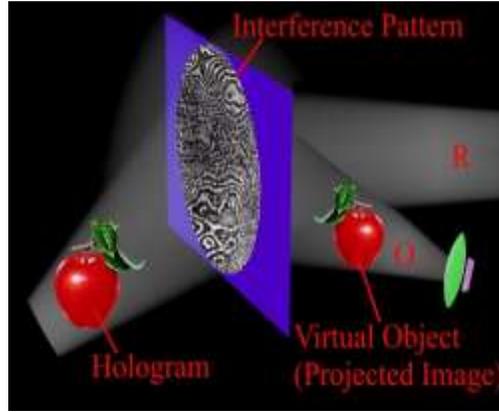

Fig. 2.The configuration of the proposed RTRH system. The light from the 3D virtual object (projected by MVS) interference with the reference light and form an interferogram on the transparent screen then at the other side of the screen a conjugate image of the virtual object reconstructed.

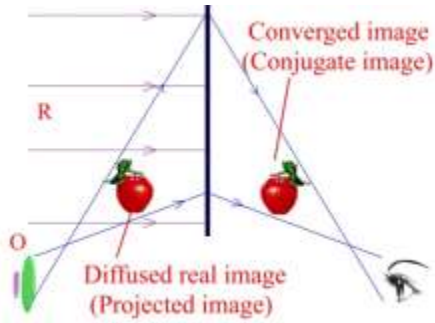

Fig. 3. Illustration of the principle of the RTRH. The RTRH can convert a divergent image (such as the image that projected by the MSV, which can't be seen completely) to a visible converged image at its conjugate position (conjugate image).

3. **ANALYSIS AND DISCUSSION**

Conventional holographic display usually writes the computer generated holographic pixel, known as a 'hogel', in a recording material (such as photorefractive polymer) and then illuminate it with the reading light to display the hologram. The update function is realized by repeat the sequence of write, read, erasure and rewrite. However, the lack of sufficient computational power to produce realistic computer-generated holograms and the absence of large-area and dynamically updatable holographic recording media have prevented realization of the concept[13]. But these problems are avoided in our system. In fact, there is no need to write the hogel physically in a material. If we can display the hogel on a screen, then the hologram can be reconstructed. Instead of writing the interferogram in a specific kind of material, in our designed system, the interferogram is generated by the direct interference of the reference light and the object light, and the interferogram is directly projected on a rear projection screens (or a transparent scatter screen). This physical process takes almost no time (no matter how complex the displayed object is). Then the writing process and erasure process are eliminated. This feature allows it to realize the function of real-time holographic. And there is no need to generate the data for the interferogram. Hence, a commonly used mobile phone processor (which can afford real-time rendering of 3D online game) is quite sufficient for the associate data processing and this in turn ensures the affordability of the system to a large extent.

One of the default defects of conventional projection system is the low brightness, which makes it almost impossible to work outside. This is mainly because of the reason that the light scattered by the screen attenuates as the inverse square of the distance from the screen. Nevertheless, this problem is eradicated in the novel 3D display. The RTRH process makes sure that the diffractive light will converge at the watch window. Another advantage of the RTRH is that the generated holographic interferogram gets rid of the constraints of the resolution of photographic film. This in turn helps a lot in improving the quality of the image. In traditional hologram technology, the image quality is easily affected by speckles that generated by the random interference among object points[4]. Also, the recording condition of a conventional hologram is really strict, even the vibration caused by talking (or thermal expansion）may cause the failure of the recording process. Meanwhile, chromatic hologram is also hard to realize in traditional hologram display. All of these inherent disadvantages of tradition holographic technique have been addressed in the RTRH technology. And the 3D image generation process of this novel 3D display is just like extruding 2D pictures into a 3D one.

The viewing space is another important feature of a 3D display. The viewing space of the proposed novel 3D display depends on the MSV. The working principle of MSV is similar to a projector or camera. By adjusting the focal plane of projector (or camera) we can project picture (or take photos) at different depth. Similarly, the MSV can also project image at different depth through the adjusting of focal plane. Like a projector, the MSV has image chip (LCD or DMD) and projection lens, usually a little change of the distance between image chip (LCD or DMD) and projection lens may cause a large change of display focal plane—this is the reason why it called micro-volumetric scanning system. Then if properly designed, we can obtain a viewing depth of sever meters or even several hundred meters while the scanning depth of MSV is about millimeter-level. Take mobile phones as an example. Even the moving distance of the lens in a cellphone camera is much smaller than a centimeter (the thickness of a cell phone is only about 1cm), the camera can take photos clearly at a distance from one tenth meter to several meters. According to the reversibility of optical path, it means that using a similar system we can acquire a display volume of about several cubic meters even the oscillation amplitude of moving parts (image generation chip) is much less than a centimeter. What we need to do is to replace the CCD of a cellphone camera with an image generation chip, such as a digital mirror device (DMD) or some other things with the similar function. In order to realize a proper display space we need to configure the system reasonably. Another benefit of the novel 3D display system is that the visible solid angle can be very large. The visible solid angle is the solid angle that is determined by the boundary of the screen and the watch widow ($\sim \frac{\text{the area of the screen}}{\text{distance from the screen to watch window}}$). Any image point inside the solid angle, no matter it is in front of the screen or behind the screen, is visible. The feeling of using such a 3D display is similar to that watching the real scenery through a window. So, by using larger screen, the larger visible solid angle can be obtained. This feature makes sure that enlarging the image's spatial does not increase the cost of the system obviously. For example, a screen of about one square meter with a watch distance of one meter having solid angle around 1 can be easily achieved. In addition, even the viewpoint is limited in the above designed system, there are a number of ways to extend it to a multi-view one.

## 4. CONCLUSION

The present study proposes a novel 3D display contains a MVS and a RTRH. The MVS can project a series of 2D images at different depth of field, which can form a 3D image in true 3D space. The RTRH is an improvement of traditional hologram system. The projected 3D image (by the MVS) works as a virtual object in the RTRH. And the interference pattern of virtual object light and reference light is displayed on a rear projection screens. Thus, instead of illuminating the interference pattern recorded in a recording material, it directly shows the lighted interference pattern (of virtual object light and reference light) on a transparent projection screen. Then the hologram of the virtual object is reconstructed timely. This special design allows real time refresh of hologram and provide higher quality images, large viewing space and more advanced features. All of the inherent disadvantages of tradition holographic technique have been addressed in the RTRH technology. So it's the most promising commercial available 3D display technology.